\documentclass[twocolumn]{aastex631}

\newcommand{\modot}{$\dot{\text{M}}_{\odot}$}
\newcommand{\msun}{$\text{M}_{\odot}$}

\accepted{November 2, 2021}

\submitjournal{ApJ}

\usepackage{savesym}
\savesymbol{tablenum}
\usepackage{siunitx}
\restoresymbol{SIX}{tablenum}
\shorttitle{CMEs and type II radio emission on $\epsilon$ Eridani}
\shortauthors{\'{O} Fionnag\'{a}in et al.}
\graphicspath{{./}{figures/}}

\begin{document}

\title{Coronal mass ejections and type II radio emission variability during a magnetic cycle on the solar-type star $\epsilon$ Eridani}

\correspondingauthor{D\'{u}alta \'{O} Fionnag\'{a}in}
\email{dualta.ofionnagain@nuigalway.ie}
\author[0000-0001-9747-3573]{D\'{u}alta \'{O} Fionnag\'{a}in}
\affiliation{{\rm Centre for Astronomy, School of Mathematical and Statistical Sciences,
National University of Ireland Galway, 
Galway, Ireland}}
\author[0000-0002-1486-7188]{Robert D. Kavanagh}
\affiliation{{\rm School of Physics, Trinity College Dublin, The University of Dublin, Dublin 2, Ireland}}
\affiliation{{\rm Leiden Observatory, Leiden University, PO Box 9513, 2300 RA Leiden, The Netherlands}}
\author[0000-0001-5371-2675]{Aline A. Vidotto}
\affiliation{{\rm School of Physics, Trinity College Dublin, The University of Dublin, Dublin 2, Ireland}}
\affiliation{{\rm Leiden Observatory, Leiden University, PO Box 9513, 2300 RA Leiden, The Netherlands}}
\author{Sandra V. Jeffers}
\affiliation{{\rm Max Planck Institute for Solar System Research, Justus-von-Liebig-Weg 3, 37077 Göttingen, Germany}}
\author[0000-0001-7624-9222]{Pascal Petit}
\affiliation{{\rm IRAP, Université de Toulouse, CNRS, UPS, CNES, 14 Avenue Edouard Belin, 31400, Toulouse, France}}
\author[0000-0001-5522-8887]{Stephen Marsden}
\affiliation{{\rm Centre for Astrophysics, University of Southern Queensland, Toowoomba, Queensland 4350, Australia}}
\author[0000-0002-4996-6901]{Julien Morin}
\affiliation{{\rm LUPM, Université de Montpellier, CNRS, Place Eugène Bataillon, F-34095 Montpellier, France}}
\author[0000-0001-8208-4292]{Aaron Golden}
\affiliation{{\rm Centre for Astronomy, School of Mathematical and Statistical Sciences,
National University of Ireland Galway, Galway, Ireland}}
\affiliation{{\rm Armagh Observatory and Planetarium, College Hill, Armagh, BT61 9DB, United Kingdom}}
\collaboration{8}{(BCool Collaboration)}
\collaboration{0}{\vbox to
0pt{\centerline{---}}}
%
%
%
\begin{abstract}
We simulate possible stellar coronal mass ejection (CME) scenarios over the magnetic cycle of $\epsilon$ Eridani (18~Eridani; HD 22049). We use three separate epochs from 2008, 2011, and 2013, and estimate the radio emission frequencies associated with these events. These stellar eruptions have proven to be elusive, although a promising approach to detect and characterise these phenomena are low-frequency radio observations of potential type II bursts as CME induced shocks propagate through the stellar corona. Stellar type II radio bursts are expected to emit below 450 MHz, similarly to their solar counterparts. We show that the length of time these events remain above the ionospheric cutoff is not necessarily dependent on the stellar magnetic cycle, but more on the eruption location relative to the stellar magnetic field. We find that these type II bursts would remain within the frequency range of LOFAR for a maximum of 20-30 minutes post-eruption for the polar CMEs,  (50 minutes for 2\textsuperscript{nd} harmonics). We find evidence of slower equatorial CMEs, which result in slightly longer observable windows for the 2008 and 2013 simulations. Stellar magnetic geometry and strength has a significant effect on the detectability of these events. We place the CMEs in the context of the stellar mass-loss rate (27--48 $\times$ solar mass-loss rate), showing that they can amount to 3-50\% of the stellar wind mass-loss rate for $\epsilon$ Eridani. Continuous monitoring of likely stellar CME candidates with low-frequency radio telescopes will be required to detect these transient events. 
\end{abstract}



\section{Introduction} \label{sec:intro}

$\epsilon$ Eridani is a solar-type K2V star, with a mass of $0.82\pm0.05$ \msun  \citep{Baines2012}. It is an interesting target for many reasons; it is a young star with an age between $400$ to $800$ Myr (\citealt{Barnes2007}, although it's age is largely uncertain, see \citealt{Janson2008}); millimetre observations with SMA (Submillimeter Array) and ALMA (Atacama Large Millimeter Array) have shown it possesses a debris disk \citep{MacGregor2015,Booth2017}; it displays millimetre flares \citep{Burton2021}; at least one exoplanet orbits the star ($M_p = 0.78^{+0.38}_{-0.12}~M_{\rm Jup}$, $a = 3.48 \pm 0.02$~au, \citealt{Hatzes2000,Mawet2019}); and it may be a source of $\gamma$-rays \citep{Riley2019}. It is the closest solar-type single star to the Sun (10\textsuperscript{th} closest stellar system overall) at a distance of $3.220\pm0.001$ pc \citep{gaia2021}. It has been observed many times in the GHz radio regime, with ambiguous detections at multiple frequencies (see Section \ref{sec:radio}), and observed at lower frequencies (157 MHz) with no detections using the GMRT (Giant Metrewave Radio Telescope, \citealt{George2007}). Table \ref{tab:stellar} lists the pertinent stellar parameters for $\epsilon$ Eri.

\citet{Gray1995} discovered magnetic variability in the form of Ca \textsc{ii} H \& K S-indices that suggested a 5-year magnetic cycle. This was re-examined by \citet{Metcalfe2013} who used over 40 years of Ca \textsc{ii} H \& K observations from a range of observatories. The authors found evidence for coexisting coherent cycles of 2.95 and 12.7 years followed by a low-activity minimum during the 1980's for 7 years which was superseded by the 3 year cycle once again. This 3-year cycle is evident in XMM-Newton X-ray spectra of the star \citep{Coffaro2020}, which shows that the X-ray flux varies by a factor of 2 in observations taken from 2003-2018. This cyclic period and phase appears to be in good agreement with those of chromospheric markers. 

\citet{Jeffers2014} presented measurements of the large-scale surface magnetic field over a period of almost 7 years from January 2007 to October 2013 using Zeeman Doppler Imaging (ZDI; \citealt{Donati2009}). The authors found that there is significant variability in the axisymmetry of the large-scale magnetic field, as well as a mostly dominant poloidal component to the field (apart from 2010 when the toroidal field dominates). For a complete view of the large-scale magnetic variability see Figure 4 in \citet{Jeffers2014}. In this work we use three of these large-scale magnetic fields as boundary conditions for our simulations of the stellar wind, January 2008, October 2011, and October 2013 (see Section \ref{sec:ss-wind}). \citet{Petit2021} used a multi-wavelength approach to gain a more holistic impression of the magnetic field of $\epsilon$ Eri. Near-infrared observations with SPIRou and examination of Fe \textsc{i} line cores allowed the authors to place a constraint of 1835 G on the maximum magnetic field strength, with a filling factor of 0.13. This is somewhat larger than previous estimates of magnetic field strength through Zeeman broadening \citep{Valenti2005,Lehmann2015}. 

Magnetic variability can have significant effects on the evolution of the star by modulating its angular-momentum loss \citep{Finley2019}. Each magnetic harmonic (dipolar, quadrupolar, etc.) results in different amounts of open flux regions, which can couple to the stellar wind more effectively. \citet{Wood1998} measured the mass-loss rate of $\epsilon$ Eri by taking Lyman-$\alpha$ observations of the star with HST(GHRS) and estimating the effect of charge exchange at the stellar wind-interstellar medium boundary. They found a mass-loss rate of 30 times the solar mass-loss rate (30 \modot, \citealt{Wood2018}; where the solar mass-loss rate is taken as \modot~$= 2 \times 10^{-14}$ \msun~${\rm yr}^{-1}$). \citet{Rodriguez2019} estimated from VLA observations at 33 GHz a stellar mass-loss rate of 3300 \modot, however they could not rule out other sources of emission. \citet{Suresh2020} ruled out this wind scenario from the previous authors using radio detections in the $2\text{--}4$ GHz range but only reduced the wind mass-loss rate upper limit by a factor of 2.2 to $1500$ \modot. 

On the Sun, increased magnetic activity leads to solar coronal mass ejections (CMEs), which are large energetic eruptions of plasma and magnetic field from the lower atmosphere of the Sun into the corona and heliosphere. Free magnetic energy in closed coronal loops is converted to kinetic energy through magnetic reconnection, which launches material into the corona \citep{Vourlidas2000}. These events are usually accompanied by flaring and have associated radiation that crosses the electromagnetic spectrum, from X-rays from the reconnection site \citep{Sindhuja2020} to radio from the corona due to coherent plasma processes \citep{Melrose2017,Kumari2017,Morosan2021}. 
Not all solar flares have an associated CME, but almost all X-class flares do \citep{Yashiro2009}. Of these CMEs, roughly $40\%$ are estimated to be radio loud (i.e. detectable in radio) \citep{Kharayat2021}, although radio quiet CMEs do also exist \citep{Gopalswamy2008,Carley2020}, and make up around 40\% of fast and wide solar CMEs.

There has been some evidence for stellar CME events on other stars in the form of Doppler-shifted lines alluding to plasma material being ejected from the star. This phenomena has been observed using Balmer lines in the optical \citep{Vida2016,Vida2019}, as well as at EUV and X-ray wavelengths \citep{Leitzinger2011,Argiroffi2019}. Although in some cases the evidence is ambiguous with Doppler velocities less than the stellar escape velocities (see \citealt{Leitzinger2020} for an unabridged discussion on stellar CME detection methods). Type II radio bursts---although another promising method of stellar CME detection---have not yet yielded any positive detections \citep{Osten2015,Villadsen2017,Crosley2018,Crosley2018b}. These bursts are characterised by their slowly drifting emission bands at metric wavelengths, produced at the fundamental plasma frequency and the 2\textsuperscript{nd} harmonic. These events present a possible fruitful method of detecting stellar CMEs. \citet{Moschou2019} showed that these stellar CMEs could be many orders of magnitude larger than their solar counterparts (deduced from stellar flare X-ray energies), making them promising targets in the radio regime. 

The current lack of low-frequency detections of CMEs from stars could be due to the transient nature of these events, or due to the ionospheric cut-off frequency (around 10 MHz) blocking the radio emission from reaching ground-based observatories. $\epsilon$ Eri is a favoured target for these type II radio emissions as it is young, active, and the closest solar-type star to the Sun. \citet{AlvaradoGomez2019} showed that for M dwarf stars, CMEs can be partially or completely suppressed by the strong background magnetic fields. Additionally, the authors demonstrated that the shocked super-Alfv\'{e}nic region for escaping CMEs can be quite distant from the star, resulting in emission frequencies below the ionospheric cut-off frequency. For $\epsilon$ Eri we can expect weaker magnetic fields than these extremely active M dwarfs, hopefully resulting in more eruptive events leading to shocks and strong radio emission. 

These events could have drastic affects on their stellar environments. It is estimated that stars with high CME occurrence rates could lose a significant amount of mass and angular momentum through this mechanism \citep{Aarnio2012,Drake2013,Osten2015,Cranmer2017}. In the case of young stars, this mass-loss could amount to significantly more than the total stellar wind mass-loss rate \citep{Odert2017,Cranmer2017}. For these young stars, CMEs could contribute to the erosion of circumstellar disks \citep{Osten2013}. Orbiting exoplanets could also experience increased atmospheric mass-loss due to stellar CME activity, with the most powerful CMEs capable of removing large planetary atmospheres \citep{Segura2010,Airapetian2016,Drake2019,Hazra2021}. Ignoring the effect of the debris disk surrounding $\epsilon$ Eri \citep{Greaves1998,Booth2017}, scaling relations from \citet{Odert2017} suggest that $\epsilon$ Eri b would experience anywhere from 10 to 50 CME impacts per day (assuming a flare energy distribution with index between 1.5 and 1.8, comparable to values assumed by \citealt{Coffaro2020}). Depending on the CME duration and the stellar CME-flare association rate, this exoplanet could be enduring a constant barrage of CMEs from its host star. 

The goal of this work is to investigate the effect different magnetic fields throughout the stellar cycle of $\epsilon$ Eri has on the eruption, formation, and propagation of CMEs. This is accomplished by running 3D MHD simulations to find a solution to the stellar wind (Section \ref{sec:ss-wind}) and then causing a CME to erupt from the surface of the star, the models of which are described in Section \ref{sec:cme-model}. We characterise the CME properties and derive plausible radio frequency emission ranges from these simulations, described in Section \ref{sec:radio}. We conclude on the results of our work in Section \ref{sec:conclusions}.
\begin{deluxetable}{lcc}
\tablecaption{Stellar parameters of $\epsilon$ Eridani. \label{tab:stellar}}
\tablehead{\colhead{Parameter} & \colhead{Value} & \colhead{Reference}}
\startdata
        Sp. Type & K2V &\citet{Heiter2015}\\
        T$_{\rm eff}$ [K] & 5076 & \citet{Heiter2015}\\
        log(g) & 4.61 & \citet{Heiter2015}\\
        M$_{\star}$ [\msun] & $0.82\pm0.05$& \citet{Baines2012}\\
        R$_{\star}$ [R$_{\odot}$] & $0.74\pm0.01$& \citet{Baines2012} \\
        P$_{\rm rot}$ [d] & 10.33 & \citet{Jeffers2014}\\
        Age [Gyr] & 0.4-0.8 & \citet{Mamajek2008}\\
\enddata
\end{deluxetable}

\section{The stellar wind of $\epsilon$ Eri} \label{sec:ss-wind}
To calculate a solution for the converged stellar wind of $\epsilon$ Eri we use the \textsc{batsrus} code (Block Adaptive Tree Solar Roe-Type Upwind Scheme; \citealt{Powell1999,Sokolov2013,Gombosi2018}). Specifically, we use the \textsc{awsom} version of this code, which uses the energy stored in Alfv\'{e}n waves to heat and accelerate the stellar wind \citep{VanDerHolst2014}. The acceleration is achieved by the dissipation of turbulent Alfv\'{e}n waves in the plasma. The model includes the reflection and damping of these waves as they propagate throughout the stellar atmosphere. \textsc{awsom} includes electron heat conduction (collisional below 5 R$_{\star}$; \citealt{Spitzer1953}; and collisionless above that distance; \citealt{Hollweg1976}) and radiative cooling. Energy dissipation is partitioned into three different temperatures, two anisotropic ion temperatures (parallel and perpendicular to the magnetic field) and one isotropic electron temperature. This model has been benchmarked using observations of the solar corona \citep{Vanderholst2010,Jin2012,VanDerHolst2014}. There are many examples now of this code being adapted to stellar cases, and with a wide range of stellar wind types simulated including M dwarfs, solar-type stars, and late sub-giants \citep{Garraffo2017,AlvaradoGomez2018,AlvaradoGomez2019,BoroSaikia2020,OFionnagain2021,Kavanagh2021}. 
\begin{deluxetable}{lr}
\tablecaption{Input parameters of the \textsc{awsom} model. Values are similar to those of \citet{Chandran2011,VanDerHolst2014,BoroSaikia2020}, except for the input Alfve\'{e}n wave flux ($S_A/B$), which is scaled to give mass-loss rates consistent with those of observations \citep{Wood2018}. \label{tab:inputs}}
\tablehead{\colhead{Parameter} & \colhead{Value}}
\centerwidetable
\startdata
        n & $2 \times 10^{10}$ cm$^{-3}$ \\ 
        T$_{\rm chromo}$ & 50,000 K \\
        S$_{A}$/B & $1.3 \times 10^7$ W m$^{-2}$ T$^{-1}$\\
        L$_{\perp}$ $\sqrt{B}$ &  $1.5 \times 10^5$ m $\sqrt{T}$\\
        h$_S$ & 0.17 \\
        $\alpha$ & 1.05 \\
        r$_H$ & 5 R$_{\star}$ \\
\enddata
\end{deluxetable}

Input parameters for the simulations, shown in Table \ref{tab:inputs}, are identical to those of \citet{VanDerHolst2014} apart from the base Alfv\'{e}n wave flux. Listed inputs include the base density (n), chromospheric temperature (T$_{\rm chromo}$), wave damping length (L$_{\perp} \sqrt(B)$), stochastic heating parameter (h$_S$), heat flux parameter ($\alpha$), and collisionless heat conduction parameter (r$_H$). It has been shown that varying the base chromospheric density parameter does not produce a significant variation in the simulated wind solution \citep{Lionello2009,VanDerHolst2014}. While the base temperature parameter is used in many wind models to scale the wind velocity and strength, it does not have a significant effect on our wind solutions. This is because the pressure gradients in our simulations are primarily due to the heating of the wind through the dissipation of alfven waves, and not the coronal temperature alone. This is also the case in other research that uses the AWSoM simulations such as \citet{BoroSaikia2020}. The authors investigated how the input Alfv\'{e}n wave flux (S$_A$/B) at the chromosphere affects the final mass-loss rate of a stellar wind. The resulting trend allows us to gauge the total mass-loss of our solution. As previously stated, Ly-$\alpha$ observations of the hydrogen wall between the $\epsilon$ Eri stellar wind and the ISM, \citet{Wood1998} were able to constrain the mass-loss rate of $\epsilon$ Eri to $30~$\modot. We previously mentioned radio observations have suggested larger wind mass-loss rates, however sources are ambiguous and only provide upper limits. Therefore, we aim to reproduce the mass-loss rate from \citet{Wood1998} with our initial simulation, using the January 2008 ZDI map as it is the closest ZDI map in time to the Ly-$\alpha$ observations. To achieve this mass-loss rate we implement an input wave flux for our simulations of $1.3 \times 10^7$ W m$^{-2}$ T$^{-1}$. 
\begin{deluxetable}{lccc}
\tablecolumns{4}
\tablehead{\colhead{} & \colhead{2008 January} & \colhead{2011 October} &  \colhead{2013 October}}
\tablecaption{Large-scale ZDI map geometries from \citet{Jeffers2014}. Toroidal and poloidal relate to the percentage of total magnetic energy in those fields. Dipolar ($\ell=1$), Quadrupolar ($\ell=2$), and Octopolar ($\ell=3$) harmonics the total percentages that make up the poloidal field. Axisymmetry is a percentage of the magnetic energy in m = 0 mode. \label{tab:zdimapinfo}}
\startdata
        B$_{\rm max}$ [G]& $28\pm3$ & $32\pm1$ & $42\pm2$ \\
        B$_{\rm mean}$ [G]& $10\pm1$ & $10\pm1$ & $20\pm1$ \\
        Toroidal [\%] & $6\pm3$ & $26\pm7$ & $22\pm7$\\
        Poloidal [\%] & $94\pm3$ & $74\pm8$ & $78\pm9$\\
        Dipolar [\%$_{\rm pol}$] & $64\pm5$ & $65\pm10$ & $85\pm2$\\
        Quadrupolar [\%$_{\rm pol}$] & $23\pm1$ & $15\pm3$ & $5\pm1$\\
        Octopolar [\%$_{\rm pol}$] & $8\pm2$ & $10\pm3$ & $6\pm1$\\
        Axisymmetry [\%$_{\rm pol}$] & $58\pm12$ & $63\pm10$ & $36\pm1$\\
\enddata
\end{deluxetable}
We use this parameter for each of our simulations. Since the Ly-$\alpha$ observations were over a decade prior to the ZDI observations, this mass-loss rate is considered to be a marker and not a strict constraint, as the 3-year magnetic cycle on $\epsilon$ Eri means there could be variation between the observation dates. \par
\begin{figure}
    \centering
    \includegraphics[trim=0 30 0 0, clip, width=\columnwidth]{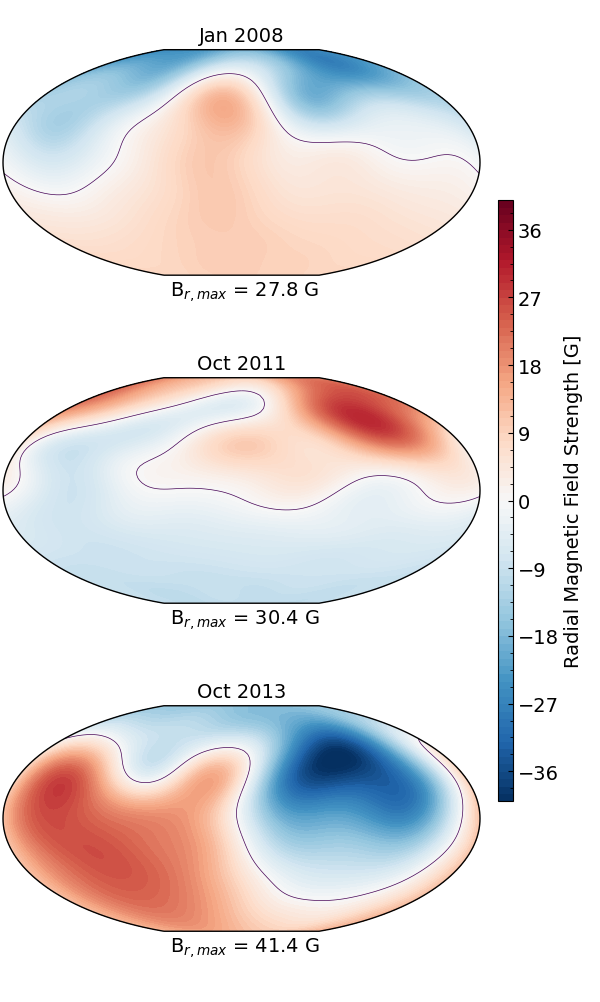}
    \caption{Surface radial magnetic field strengths from ZDI maps for the three epochs 2008, 2011, 2013 \citep{Jeffers2014}.}
    \label{fig:bfields}
\end{figure}

 
The observed magnetic field maps for $\epsilon$ Eridani for the dates January 2008, October 2011, and October 2013 are shown in Figure \ref{fig:bfields}. We chose these three epochs to simulate as they encompassed the widest range of observational dates (excluding January 2007 as it had less phase coverage), as well as providing some variation in the field axisymmetry, polarity, and strength. The magnetic field geometry values are given in Table \ref{tab:zdimapinfo}. They show that our choice of magnetic fields give a good range in maximum (B$_{\rm max}$) and mean (B$_{\rm mean}$) magnetic field strengths. Note that our MHD simulations only utilise the radial component of the measured magnetic fields.

\begin{figure}
    \centering
    \includegraphics[width=\columnwidth]{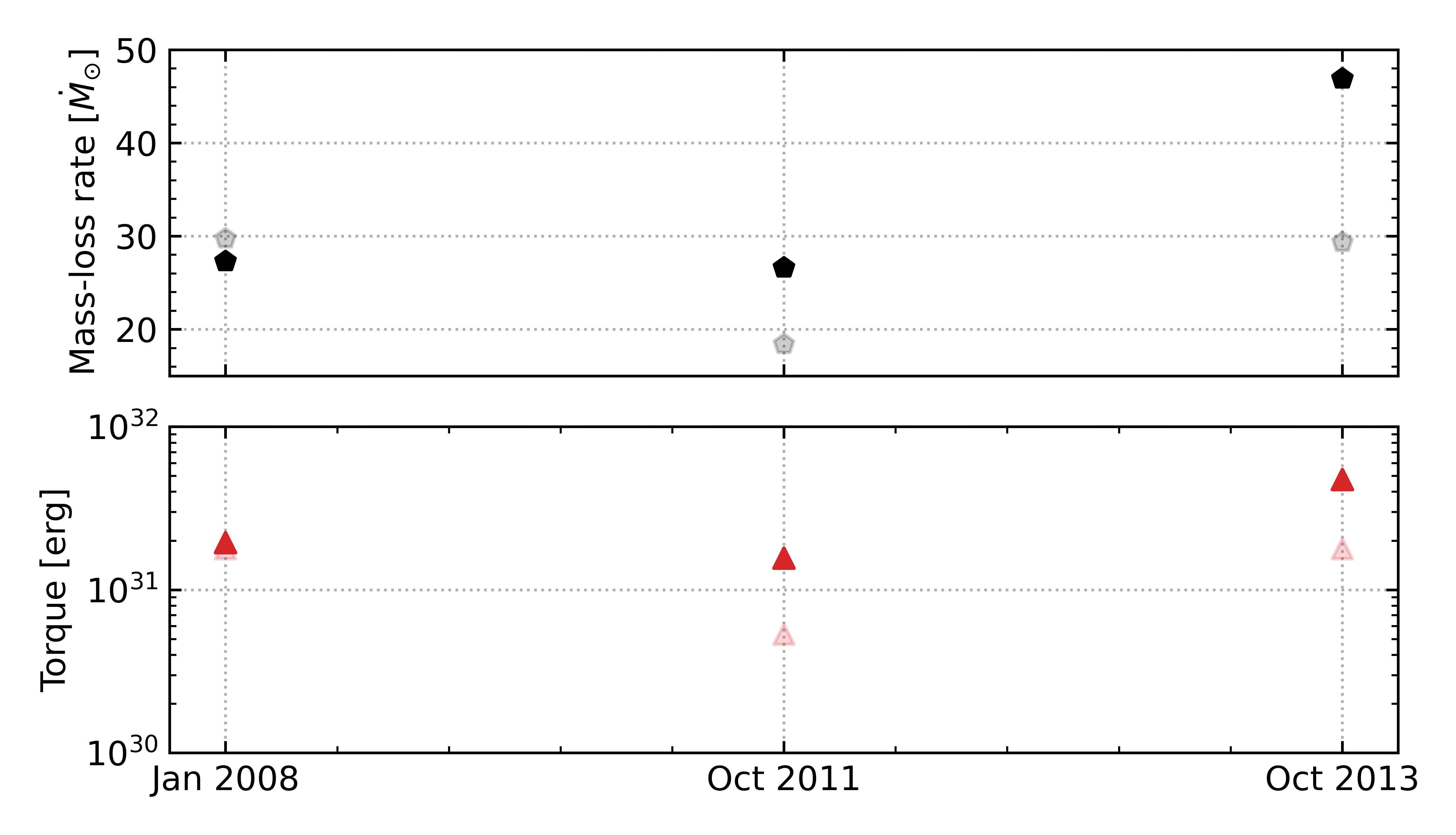}
    \caption{Mass-loss rates (black pentagons) and angular momentum-loss rates (red triangles) for each of our stellar wind simulations of $\epsilon$ Eri. \modot~is given in terms of the solar-mass loss rate (\modot~$= 2 \times 10^{-14}$ \msun yr$^{-1}$). Torque and angular momentum-loss rate are equivalent. The faded symbols are values calculated by semi-analytical relations \citep{Finley2019}.}
    \label{fig:mdotjdot}
\end{figure}
The \textsc{awsom} simulations can be prone to oscillation, but usually the solution will converge quite well with very minimal oscillation in global wind parameters such as mass-loss rate, open magnetic flux, and angular-momentum loss rate. For our purposes, we assume a simulation to be converged when these three variables oscillate below 5\%. Our mass-loss rate benchmarked January 2008 simulation is quite close to the observed values given by \citet{Wood1998}. We see this increase for the final epoch simulated, with a $\times1.7$ increase in the mass-loss rate between Jan 2008 and Oct 2013. Our simulations result in mass loss rates that are \SIrange[mode=text]{27}{48}{} times the solar mass loss rate. Angular momentum loss rates range from $1.6\times10^{31}$ erg in October 2011 to $4.7\times10^{31}$ erg in October 2013. \citet{Finley2019} used semi-analytical relations to estimate the mass-loss and angular momentum loss variability of this star. The values we find agree very well with their calculations, including the variability trend for each epoch. This is shown in Figure \ref{fig:mdotjdot}. We see excellent alignment in the 2008 epoch. In Figure \ref{fig:ss_wind} we show the velocity and magnetic field structure of the stellar wind solutions for each epoch.

\section{Modelling coronal mass ejections from $\epsilon$ Eri} \label{sec:cme-model}
\begin{figure}
    \centering
    \includegraphics[trim=5 5 5 5,clip, width=\columnwidth]{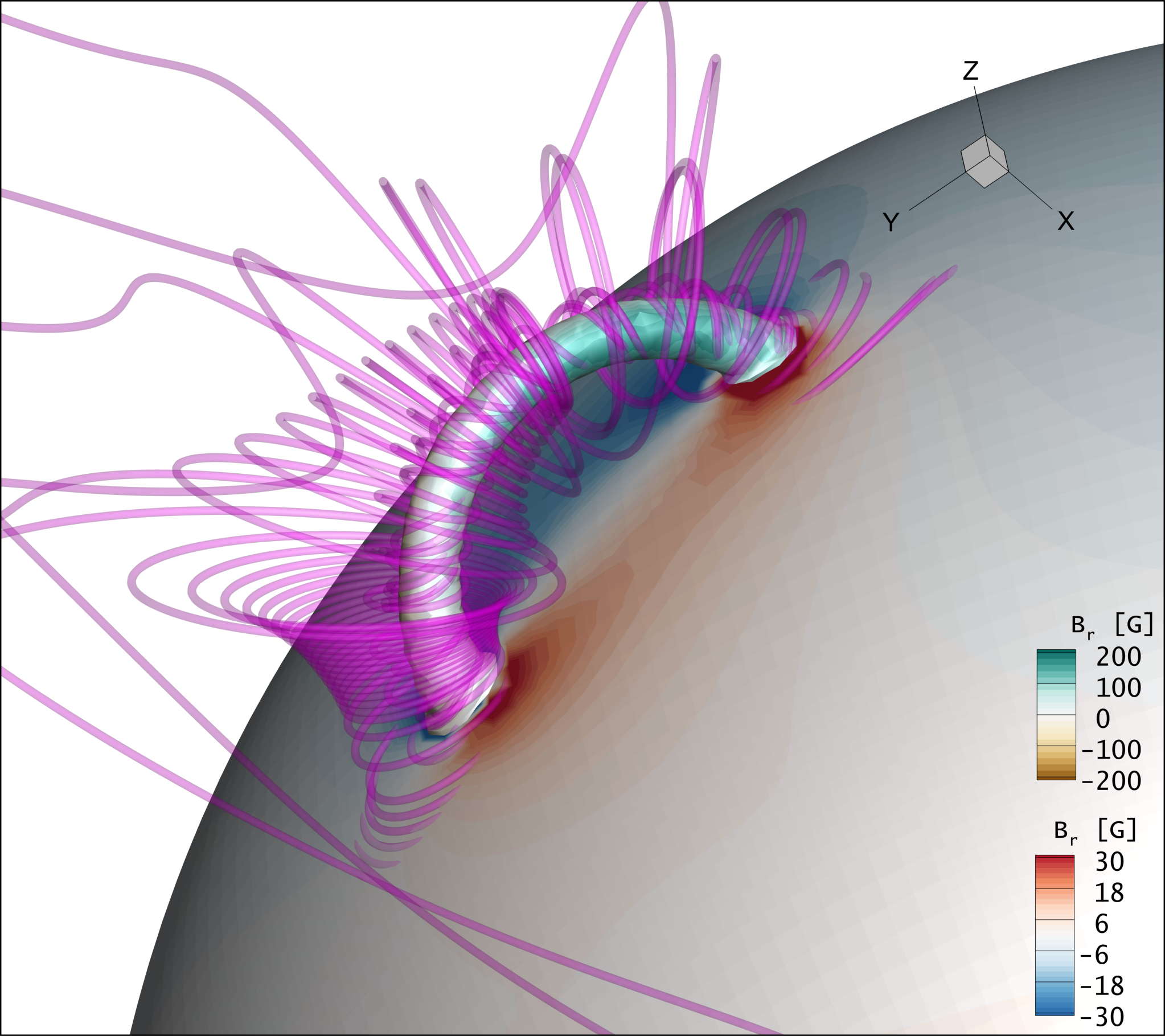}
    \caption{Image showing the initialisation of a CME for the January 2008 epoch, outlined by a high density isocontour (n = $3\times10^{8}$ cm$^{-3}$). Streamtraces showing magnetic field that interact with the flux tube are shown in magenta. Two contours are shown for magnetic field (blue-red for the stellar surface, and brown-green for the flux tube) to show the range effectively, with relatively strong magnetic fields observed around the flux tube reaching 200 G.}
    \label{fig:fluxtube}
\end{figure}

Our CME simulations are initialised using a TD flux rope model \citep{Titov1999} which is implemented within the SWMF framework using the two temperature model described by \citet{Meng2015}. This places a twisted force-free flux rope at the inner boundary of our simulations, with footpoints anchored in the stellar surface. It creates a strong, localised, magnetic bipole on the surface of the star. The sudden disequilibrium between magnetic and pressure forces causes the flux rope to erupt outwards into the corona and stellar wind. The model has eight free parameters, which control the size, shape, orientation, mass and free magnetic energy of the flux tube. These free parameters for each CME simulation carried out in this work are as follows: Longitude~(\ang{180}), Latitude~(\ang{+10}, \ang{+70}, or \ang{-70}), Orientation/tilt~(\ang{30}), major radius~(R$_{\rm maj}$; 128 Mm), minor radius~(R$_{\rm min}$; 20 Mm), depth ($\delta$; 1 Mm), current (I$_{\rm CME}$; $2 \times 10^{12}$ A), mass (M$_{\rm CME}$; $3.5 \times 10^{16}$ g). This configuration gives a loop length $\approx 802$ Mm.  For an in depth discussion on these parameters, we direct the reader to \citet{Titov1999} and \citet{Meng2015}. These parameters result in a CME flux rope such as that visualised in Figure \ref{fig:fluxtube}.
Since we do not have rigid constraints on eruption location for CMEs on a star like $\epsilon$ Eri, we simulate three CMEs for each stellar wind solution. One CME is positioned near the equator (latitude of \ang{+10}), and the other two are polar (latitude of \ang{70} and \ang{-70}). Note that the observed stellar magnetic field is less constrained in the southern hemisphere of the star due to the stellar inclination with respect to Earth. Therefore we conduct CME eruption simulations in both hemispheres to compare and contrast the effect this has on CME shock formation and propagation.

The size and current set the amount of free magnetic energy (E$_{\rm CME}$) in the flux tube. We choose these parameters so that E$_{\rm CME} = 3.5 \times 10^{33}$ erg. This magnetic energy is 3 orders of magnitude greater than those estimated for solar CMEs \citep{Vourlidas2000}, and yet still small (1000 times less) compared to estimates for some M dwarf eruptive events \citep{AlvaradoGomez2018}. The initial mass of the CME is constrained by taking the X-ray flux of flares on $\epsilon$ Eri \citep{Coffaro2020}, and comparing that to the X-ray flux - CME-mass relationship for the Sun \citep{Aarnio2011}. This results in a mass of $3.5\times10^{16} g$, large for a solar CME, but is still considered small ($10 - 10,000$ times less) compared to some estimations for CMEs on other stars \citep{Moschou2019}. Due to the elevated X-ray fluxes, stronger magnetic fields of $\epsilon$ Eri \citep{Petit2021}, and its young nature, the energy related to our CME constructed here is well within extended observational trends for a star of this kind.
\begin{deluxetable}{lccc}
\tablecolumns{4}
\tablehead{\colhead{Sim} & \colhead{u$_r$ [km s$^{-1}$]} & \colhead{M$_{\rm CME}$ [g]} &  \colhead{E$_{k}$ [erg]}}
\tablecaption{ Table showing some macro parameters of our simulated CMEs after 1 hour. Maximum radial velocity of the CME ($u_r$ [km s$^{-1}$]), mass of the CME (M$_{\rm CME}$), and kinetic energy of the CME (E$_k$) are tabulated.  Eq - Equatorial, NP - North Polar, SP - South Polar.  \label{tab:cme_out}}
\startdata
    2008 Eq & 2602 & $1.70\times 10^{16}$ & $5.75\times10^{32}$\\
    2008 NP & 2730 & $2.51\times 10^{14}$ & $9.35\times10^{30}$\\
    2008 SP & 2229 & $6.15\times 10^{15}$ & $1.53\times10^{32}$\\
    2011 Eq & 2685 & $2.78\times 10^{15}$ & $1.00\times10^{32}$\\
    2011 NP & 2993 & $5.34\times 10^{16}$ & $2.39\times10^{33}$\\
    2011 SP & 2247 & $6.04\times 10^{15}$ & $1.52\times10^{32}$\\
    2013 Eq & 1527 & $1.93\times 10^{15}$ & $2.25\times10^{31}$\\
    2013 NP & 2110 & $1.41\times 10^{15}$ & $3.13\times10^{31}$\\
    2013 SP & 2476 & $1.14\times 10^{15}$ & $3.49\times10^{31}$\\
\enddata
\end{deluxetable}
\begin{figure*}
    \centering
    \includegraphics[trim=2 2 2 2, clip, width=0.95\linewidth]{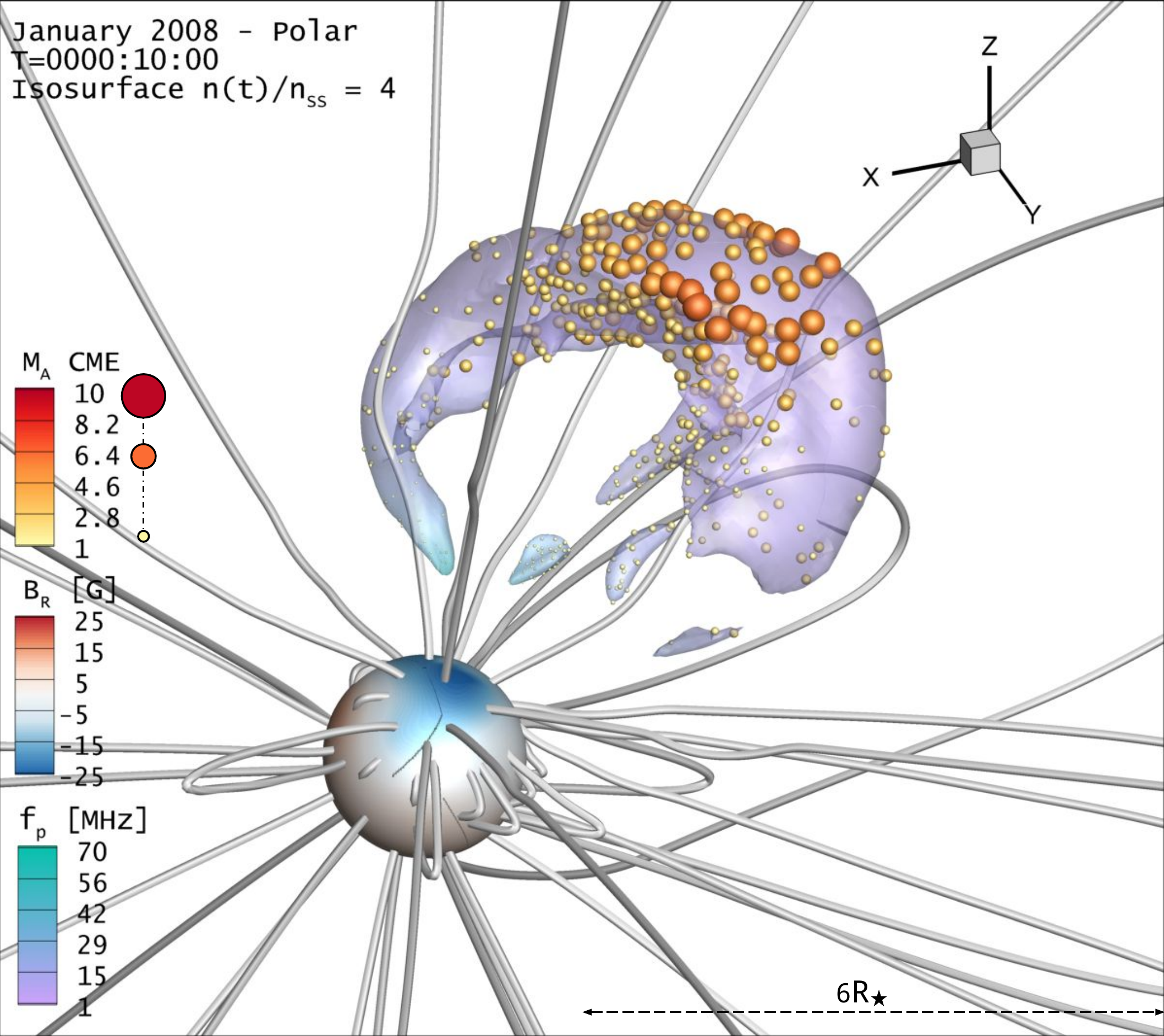}
    \caption{The CME is shown 10 minutes after eruption for the north polar January 2008 simulation. The CME is delineated by a factor of 4 difference in the plasma density ratio between this timestep (T) and the steady state wind solution (SS). The magenta-turquoise contour on the CME surface denotes the plasma frequency. The scatter is encoded with the CME Mach number value in the sphere colours (yellow-orange-red) and sizes. The red-blue contour on the stellar surface shows the magnetic field in Gauss. Magnetic field streamlines are shown in gray.}
    \label{fig:j08_eq_cme}
\end{figure*}

Figure \ref{fig:j08_eq_cme} shows an example of a simulated CME, 10 minutes post-eruption. To define the isocontour surrounding the CME volume, we compare plasma density at the current timestep to the plasma density for the steady-state stellar wind solution. A ratio of $\times4$ between these values defines the isocontour (similar to the definition in \citealt{AlvaradoGomez2019}). In reality, the CME shock and radio emission will precede the high density region of a CME by some finite distance. The shock location can alternatively be delineated by finding the maximum entropy gradient in the plasma \citep{Schmidt2013}, discussed in Appendix \ref{app:shocks}. From our simulations we calculate the distance between the CME body and entropy gradient shock location to be 0.2 R$_{\star}$. This means that the plasma frequencies calculated in this work are upper limits, with the actual value marginally lower (by a factor of 1--2) due to the shock location having a slightly lower plasma density. \par
To determine if the CME produces a shock, it is important to consider the Alfv\'{e}n Mach number,
\begin{equation}
    M_A = (u_{\rm r,sw} - u_{\rm r,cme}) / u_A,
    \label{eq:machnumber}
\end{equation}
where $u_{\rm r,sw}$ is the stellar wind velocity, $u_{\rm r,cme}$ is the CME velocity, and the Alfv\'{e}n velocity is $u_A = B/\sqrt{4\pi \rho}$. 
This CME Mach number is depicted in Figure \ref{fig:j08_eq_cme} as a scatter of spheres across the surface of the CME, where the colour and size are proportional to the Alfv\'{e}nic Mach number. From solar studies, it appears that type II radio bursts are most likely to originate from regions of high Alfv\'{e}nic Mach number, or low Alfv\'{e}n velocity, and quasi-perpendicular geometries \citep{Zucca2014,Zucca2018,Maguire2020}.
\begin{figure*}
    \centering
    \includegraphics[trim=0 25 0 0, clip, width=0.95\linewidth, page=2]{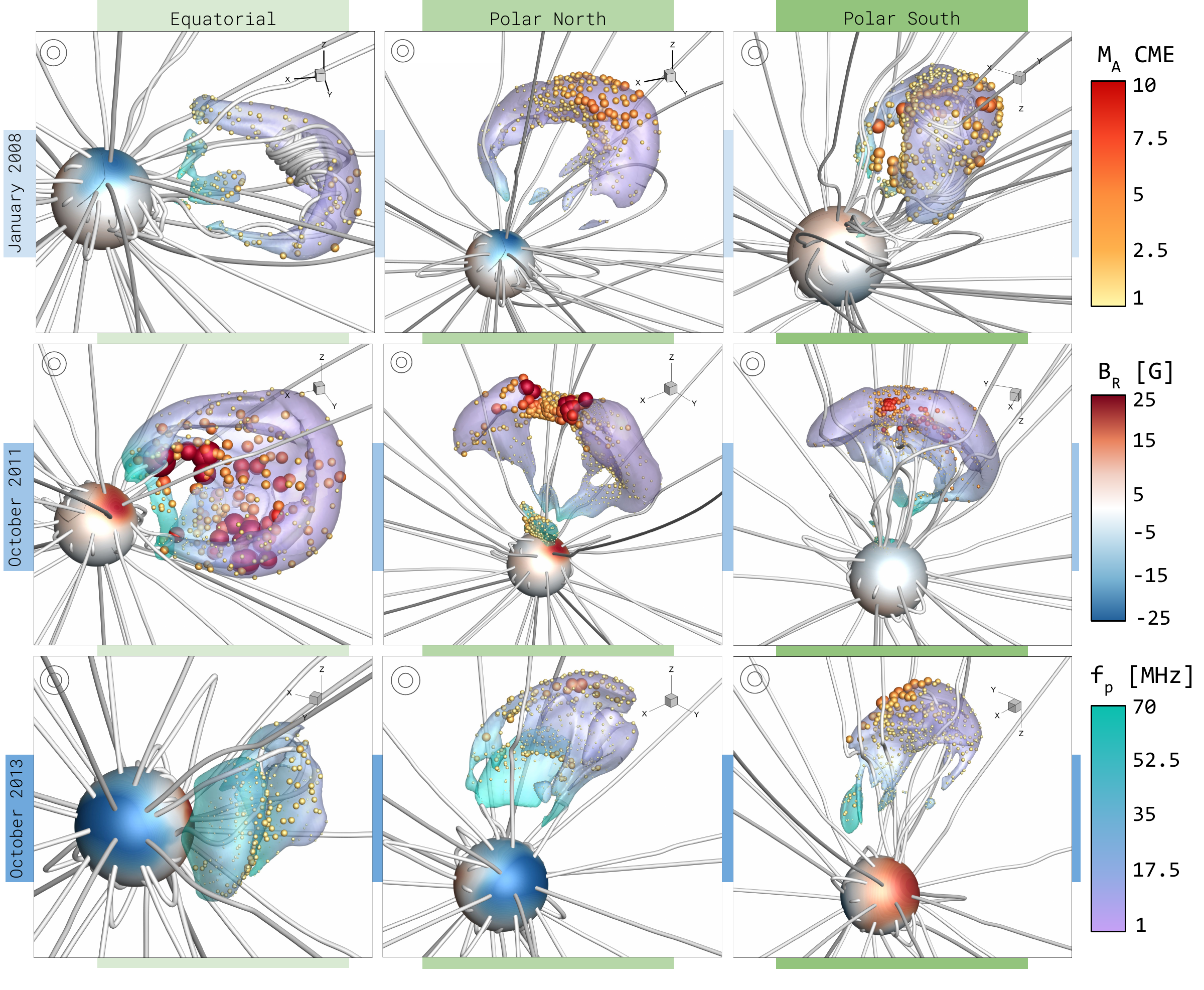}
    \caption{The nine simulations in this work are shown in this figure, both equatorial CMEs (left) and polar CMEs (middle, right), with the same style as Figure \ref{fig:j08_eq_cme}. Each row represents a different observed ZDI epoch. Each panel here shows the CME solution at 10 minutes post-eruption. The relative scatter size for $M_A = 5$ (small) and $M_A = 10$ (large) are shown in the top left of each panel.}
    \label{fig:all_cmes_10min}
\end{figure*}
The 2013 equatorial epoch presented the slowest CME velocities from our simulated sample (see Table \ref{tab:cme_out} and Figure \ref{fig:cme_vel} in Appendix \ref{app:v_dists}). This is expected as the strongest azimuthal magnetic fields are present in this case. This strong azimuthal field partially impedes the CME, providing stronger magnetic pressure resisting expansion. For this reason, at the 10 minute mark, Figure \ref{fig:all_cmes_10min} shows the equatorial CMEs as closest to the stellar surface, particularly so in the case of October 2013, which has the strongest surface magnetic field of all epochs. Table \ref{tab:cme_out} shows the CME velocity, total mass at 1 hour, and kinetic energy of each CME shown in Figure \ref{fig:all_cmes_10min}. The CME mass here is defined as the mass within the isocontours defined above, with velocities greater than the escape velocity ($v_{\rm esc} = \sqrt{2GM}/r$) of the star. Note that only one CME has a mass greater than the initial flux rope mass, which is the 2011 northern polar CME. This can occur as the CME sweeps up extra material and carries it outwards. The other cases show a reduced mass from the initial flux rope mass, which can be explained by material being trapped close to the stellar surface, or slowing below the requires escape velocity. If these CMEs occurred at a rate near the upper end of the relation described by \citet{Odert2017} (Figure 2 within; assuming $\alpha$ between 1.5 and 1.8), we could expect these CMEs to carry away $1.53\times10^{-15} - 3.25\times10^{-13}$ M$_{\odot}$ yr$^{-1}$ ($50 d^{-1}$). The lower range of this estimate amounts to $\approx 25\%$ of the stellar wind mass-loss rate for our January 2008 wind simulation, or as high as $\approx 53\%$. This is a significant proportion of the stellar mass carried away in eruptive events. At the lower end of the scale of the CME frequency relation from \citet{Odert2017}, we would still expect $2.5\% - 5\%$ of the stellar mass-loss to occur due to CMEs. Note that we assume a uniform CME-mass relation from our simulations here, not the CME-mass distribution in \citet{Odert2017}.\par 
Comparing the northern and southern hemisphere CMEs shows no clear trend between the two cases. We see a difference in eruption velocity (Table \ref{tab:cme_out}) for each scenario. While in 2008 and 2011 the northern CMEs display an average velocity of 624 km s$^{-1}$ faster than their southern counterparts, the 2013 shows the reverse scenario, with the northern CME being 366 km s$^{-1}$ slower than the southern CME. This lack of a trend is evident in CME kinetic energy where the northern event has the largest value in the 2011, but the reverse is true in the 2008 and 2013 epochs. Given that only the northern hemisphere of the stellar surface magnetic field was constrained through ZDI, it is surprising that we do not see more pronounced differences in the opposing CME events.
\section{CME Radio Emission and variability} \label{sec:radio}
While M dwarfs have been the focus of possible type II radio bursts sources for some time \citep{Villadsen2017,Crosley2018,Crosley2018b}, recent numerical studies have suggested that the coronal and wind environment around these vibrant magnetic stars is not conducive to strong type II bursts \citep{AlvaradoGomez2018,AlvaradoGomez2019}. In some cases a strong azimuthal magnetic field can cause complete suppression of an erupting CME resulting in a failed eruption, and in others there are partial eruptions (or impeded in some way, similar to our October 2013 CME simulation above). Additionally, the fast stellar winds of these stars (e.g. \citealt{Kavanagh2021}) means the CMEs only become super-Alfv\'{e}nic at large distances, meaning the density is greatly reduced, along with the plasma frequency, compared to the lower stellar atmosphere making detection more difficult \citep{AlvaradoGomez2020}. Solar-type stars such as $\epsilon$ Eri present an attractive alternative to the much more magnetic M dwarfs. As a younger and more active star, it displays frequent flaring \citep{Audard2000,Coffaro2020} combined with a background magnetic field whose strength is insufficient to confine any putative CME eruption, which we see in our results in Figure \ref{fig:all_cmes_10min}. 

Since the type II emission mechanism is coherent plasma emission due to Langmuir wave growth (see \citealt{Melrose2017} for an in depth discussion on astrophysical plasma emission mechanisms), we expect any emitted radio wave to occur near the plasma frequency ($f_p = 9000 \sqrt{n_e [{\rm cm^{-3}}]}$ Hz), in this work we assume the emission occurs at the plasma frequency. In reality, the emission would need to occur above the local plasma frequency in order to propagate without absorption. \citet{Schmidt2013} use the BAT-R-US code to simulate a solar eruptive CME event, which they combine with their stochastic growth theory for type II radio burst emission and successfully replicate the type II burst observed with the WIND spacecraft on 2011 February 15 UTC. In our simulations, we extract the plasma frequency information from the volume of the CME front at each timestep and use this to create pseudo-spectra, which consist of the frequency of the emission, but not the intensity.

$\epsilon$ Eri has been observed many times at various radio frequencies \citep{Bastian2018,Rodriguez2019,Suresh2020}. Each of these works has achieved detections of $\epsilon$ Eri at frequencies above 1 GHz. Figure 4 in \citet{Suresh2020} collates these observations into a radio spectrum of the star above 2 GHz, including millimetre observations \citep{MacGregor2015,Booth2017,Lestrade2015,Chavez-Dagostino2016}. While these observations are interesting for other reasons, such as the stellar wind constraint provided by \citet{Suresh2020}, the transient type II radio bursts occur at much lower frequencies (420 MHz maximum, 110 MHz average, \citealt{Umuhire2021}). \citet{George2007} observed $\epsilon$ Eri at 157 MHz using the GMRT, but did not detect anything. They were able to place upper limits of 7.8 mJy on the radio emission from the system. These observations were aiming to detect exoplanetary radio emission, only amounted to 4 hours on source, and were significantly affected by radio-frequency interference. To date there have been no detections of $\epsilon$ Eri at this low frequency range.

\begin{figure*}
    \centering
    \includegraphics[trim=0 10 0 0,clip, width=0.5\textwidth, page=1]{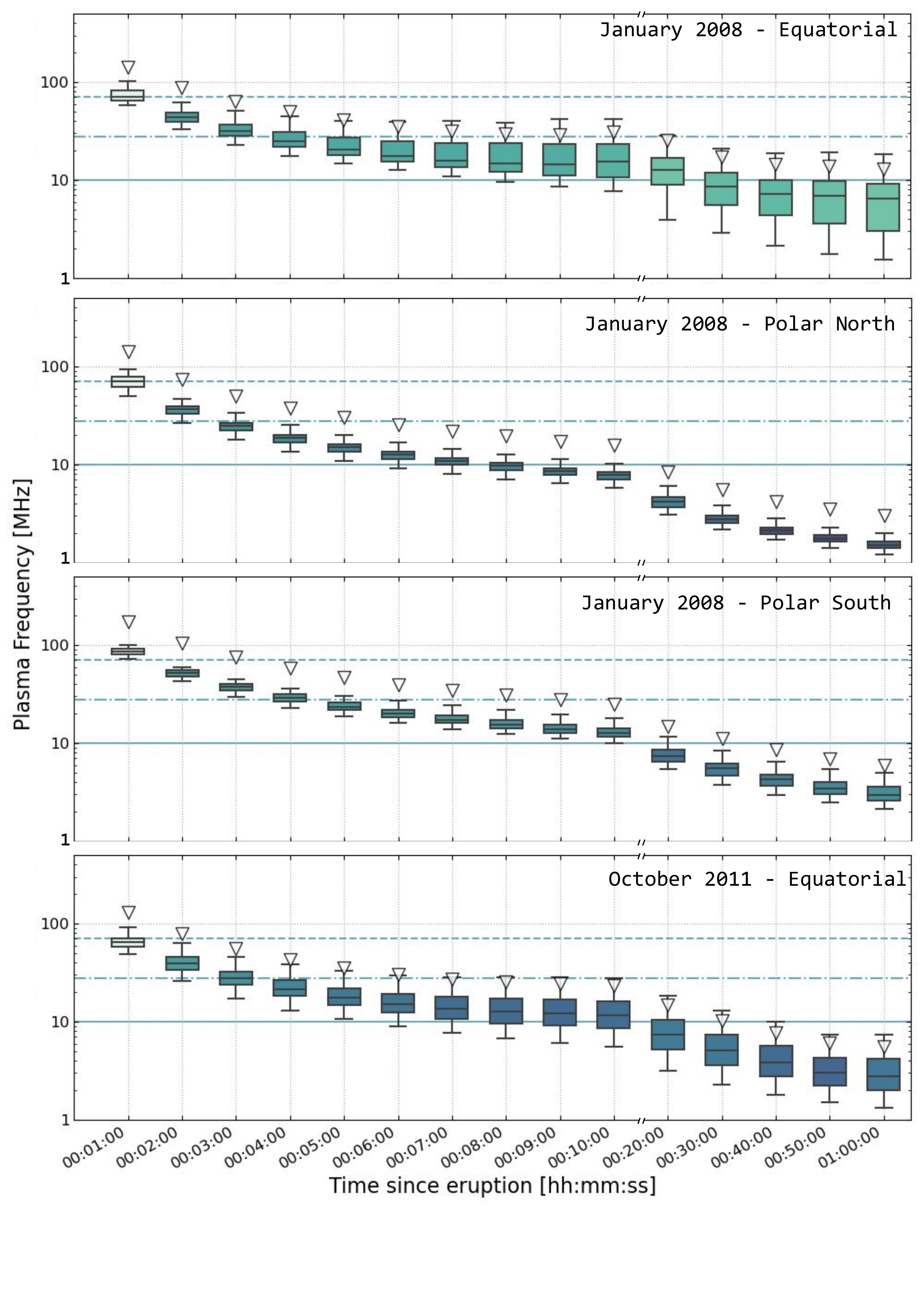}\hfill
    \includegraphics[trim=0 10 0 0,clip, width=0.5\textwidth, page=2]{spectra_CMEs.pdf}\hfill \vspace{-.8cm}
    \includegraphics[trim=0 610 0 0,clip, width=0.5\textwidth, page=3]{spectra_CMEs.pdf}
    \includegraphics[trim=180 390 120 250,clip, width=0.3\textwidth, page=3]{spectra_CMEs.pdf}\hfill
    \caption{Nine plots showing the pseudo-spectra for each of the CME events simulated in this work. The boxplots show the median and interquartile range (IQR) for each timestep extracted from our simulation. Whiskers display $1.5\times$IQR. Extreme outliers usually denoted as fliers are removed from this plot for clarity. Note the discontinuous time axis after 10 minutes. The white nabla marker denotes the 2\textsuperscript{nd} harmonic of the median in the boxplot. We show the lowest frequency attainable for LOFAR/NenuFAR (solid), LWA (dash dot), and MWA (dashed). Note frequencies here could be overestimated by almost a factor of 2 (see Appendix \ref{app:shocks}.}
    \label{fig:all_spectra}
\end{figure*}
Pseudo-spectra are shown in Figure \ref{fig:all_spectra} and depict the range of possible emission frequencies from each CME at each timestep in our simulations. The boxes show the median plasma frequency of the CME (horizontal black lines), extracted from the isosurfaces of the CMEs shown in Figure \ref{fig:all_cmes_10min}, and the inter-quartile range for the CME volume (25\% -- 75\% of the distribution). The white nabla symbols mark the 2\textsuperscript{nd} harmonic, which is seen to have strong emission for solar CMEs. Typically, outliers are denoted as fliers in boxplots, however we have removed them as they render Figure \ref{fig:all_spectra} unreadable. Outliers in this case are due to both high and low density artefacts in the CME isosurface definition and pockets of high density material appearing near the stellar surface. This results in many thousands of fliers due to the high resolution of our simulation grid. Note that due to the shock definition, it is difficult to determine exact emission frequencies without rigorous follow-up observations. For this reason, frequencies in Figure \ref{fig:all_spectra} provide an upper-limit and could be overestimated by a factor of 2 (see Appendix \ref{app:shocks}). Depending on the magnetic field epoch, and the location of eruption, the window of opportunity for a positive detection of these events is usually 10 minutes for fundamental emission using LOFAR (LOw Frequency ARray). However, variation in the October 2011 magnetic field (Figure \ref{fig:all_spectra}, top right panel) shows extended fundamental emission up to 30 minutes above the frequency range of LOFAR. Harmonic emission retains a frequency above 10 MHz for an additional 10 minutes for most cases. These times are much shorter for the other telescope frequency ranges, with fundamental emission only remaining above this level for 1-5 minutes for the MWA (Murchison Widefield Array) and LWA (Long Wavelength Array). Note that the emission is expected to exceed 150 MHz for only a short time ($\sim$1 min) and in some cases it is only the harmonic emission that does this. This, combined with sensitivity issues of the instrument at these short timescales and RFI, could explain the non-detection of any emission by \citet{George2007}. It is important to note that these pseudo-spectra do not give any information on the emission intensity, and so we can not say if emission would be strong enough to actually be detected by each of these instruments. This emission could change given more precise particle acceleration and radiative transfer descriptions. While we do not carry out this analysis in this work, the simulations presented here could be used to apply a more rigorous analysis on the radiation generated by electron beams at the CME shock location (e.g. using quasi-linear theory, \citealt{Reid2015}; or stochastic growth theory, \citealt{Knock2001,Schmidt2012a,Schmidt2012b}). \par 
On the Sun, the acceleration sites in type II related CMEs seem to vary. \citet{Zucca2018} estimated one type II event on the Sun originated from plasma on the CME \emph{flank} with Mach numbers in the range of 1.4--1.6. \citet{Maguire2020} found a solar type II radio burst generated from the CME \emph{nose} with similar Mach numbers, which ceases as the CME progresses, despite increasing Mach number. The exact location of bursts in the radio is further complicated by strong refraction in the solar corona. This makes it difficult to accurately locate solar radio bursts (see \citealt{Kontar2019} for a discussion on solar coronal scattering, and \citealt{Chrysaphi2018,Chen2020} for detailed type II and III radio burst scattering analysis respectively). \citet{Morosan2021} pointed out that slower solar CMEs ($<$ 200 km s$^{-1}$) are more likely to be associated with type II bursts, while faster CMEs are likely associated with type IV bursts. From Figure \ref{fig:all_cmes_10min}, the range of Alfv\'{e}nic Mach numbers in the plasma are shown as scattered spheres. They reach much higher values than the ones cited previously for solar CME events. However, it is clear from the October 2011 simulations that the largest Mach numbers can appear at any location around the CME surface. For the equatorial case we see high Alfv\'{e}n Mach numbers in the CME flanks, however for the polar cases the highest Alfv\'{e}n Mach numbers appear on the nose. The location of the strongest beamed electrons that would drive a type II event on these CMEs is also dependent on the orientation of the shock front and plasma velocity components to the ambient magnetic field \citep{Yuan2008,Schmidt2012b}. See Appendix \ref{app:v_dists} for examples of velocity component distributions of our CME events. 

The simulated October 2011 equatorial CME presents the highest Alfv\'{e}n Mach numbers of any of our CMEs (also evident in Figure \ref{fig:all_spectra}). In the case of this eruption, the volume it erupted into was particularly low density in the steady wind solution. This gives rise to a large defined CME structure which can produce very high Alfv\'{e}n velocities due to the low density in the wind previously.

Keeping the caveats of the pseudo-spectra in mind, i.e. uncertainty in radio burst location along the CME front and radio intensity, it remains likely that the maximum time these events on $\epsilon$ Eri would remain within the spectral coverage is 20-30 minutes post-eruption for the polar CMEs, with a slight chance of harmonic detection for up to 50 minutes.
If we compare the differences between the equatorial and polar spectra, we find that equatorial CMEs tend to exhibit much higher emission frequencies for a longer period of time post-eruption. This is likely due to their slower nature, being confined by more restrictive closed-magnetic field lines. This means they are embedded in a higher density plasma for a longer time. It is clear that the magnetic geometry and strength has a strong affect on the detectability of these events. In the synthetic spectra for the polar cases, we see a much smaller range and a faster decline of the median plasma frequency within the CME. This suggests that polar CMEs on $\epsilon$ Eri could have a shorter detectable window than their lower latitude counterparts. These effects are evident in Figure \ref{fig:all_cmes_10min}, where we see the left column of equatorial CMEs exhibit much stronger flank formations, whereas the polar CMEs retain a more globular shape. The exception is the October 2011 northern polar CME which retains a dense globular formation near the stellar surface which might skew the pseudo-spectrum towards higher plasma frequencies. It is important to remember that these pseudo-spectra are just a guide for expected emission frequencies, and actual emission frequency could differ depending on the location of strongest shock formation.

The October 2013 equatorial CME experiences a serious fragmentation after 10 minutes, and is the slowest of the CMEs. For this reason it could be considered to be partially suppressed by the stellar magnetic field. As this epoch has the strongest magnetic field of those considered in this work, it makes sense that it is the only one for which we see partial suppression. 

Of course, much more massive events, with larger flux rope masses, would likely give rise to higher plasma frequencies surrounding the shock, making them detectable for a longer period of time. However, in our simulations, our flux rope mass of $3.5 \times 10^{16}$ g is at the larger end for solar CMEs. Given that $\epsilon$ Eri is a solar-type star, we proposed this flux rope mass was a good approximation of expected values. \citet{Moschou2019} estimated that some active stars could exhibit CMEs with masses 5-6 orders of magnitude larger than the value we used in this work. These super-CMEs are more likely to exist on the most active of M dwarfs due to their strong stellar magnetic fields and intense flaring activity. 

\section{Conclusions} \label{sec:conclusions}
We simulated nine different CMEs for $\epsilon$ Eridani across three different epochs. From our assumed CME setup, each case erupted violently into the surrounding stellar wind and produced a shock. These shocks were estimated to produce type II burst radio emission as high as 165 MHz at the fundamental frequencies, and twice this value for their 2\textsuperscript{nd} harmonics. The duration that these events would remain observable was 10-20 minutes in most cases assuming an ionospheric cutoff of 10 MHz. In some particular cases, this time-frame was slightly extended to $\approx 30$ minutes (October 2011 - Polar North CME). 

We found little clear evidence of the magnetic cycle of $\epsilon$ Eridani affecting the CMEs across the epochs January 2008 to October 2013. Each epoch produced strong shocked material that was ejected from the stellar surface in each eruptive location. The lack of magnetic confinement and large Alfv\'{e}n Mach numbers are promising signs of type II radio emission from these events. In the case of the 2013 epoch, we see the equatorial CME is significantly slower than other equatorial eruptions at different epochs. This is the only evidence of magnetic variability affecting the CMEs, which results in lower Alfv\'{e}n Mach numbers, but a longer duration embedded in high density plasma.  Eruption location on the stellar surface has an effect on the CME propagation, closely tied to magnetic geometry. We found CMEs that erupt through regions of closed magnetic field are slightly impeded and exhibit slower speeds and lower Alfv\'{e}n Mach numbers, indicative of weaker shocks. This could result in much weaker type II bursts, or perhaps no formation of type II emission at all. However, if type II emission occurs, it would result in longer observable windows from low-frequency telescopes.

In the future, these CME simulations on a nearby K dwarf star could be used as inputs to more rigorous particle and radiative simulations of electron beams produced by these shocks. Particle-in-cell simulations would give precise electron distributions and radiative models such as quasi-linear simulations or stochastic growth theory could calculate emitted flux density from the electron beams. This type of work could place tight constraints on the flux densities expected from typical CME events on $\epsilon$ Eridani, and whether they should be detectable from the ground with current (i.e. LOFAR, LWA, MWA) or future (e.g. SKA-Low) radio telescopes. 

\begin{acknowledgments}
D\'{O}F would like to acknowledge funding made available through the Irish Research Council (IRC) Government of Ireland (GOI) Postdoctoral Fellowship Programme.
RDK acknowledges funding received from the IRC through the GOI Postgraduate Scholarship Programme. AAV acknowledges funding from the European Research Council (ERC) under the European Union’s Horizon 2020 research and innovation programme (grant agreement No 817540, ASTROFLOW). The authors would like to thank E. Carley for providing useful discussion concerning type II radio bursts and the underlying mechanisms. This work was carried out using the \textsc{swmf/batsrus} tools developed at the University of Michigan Center for Space Environment Modelling (CSEM) and made available through the NASA Community Coordinated Modeling Center (CCMC). The authors wish to acknowledge the Irish Centre for High-End Computing (ICHEC) for the provision of computational facilities and support. NUI Galway's Office of the Vice President for Equality and Diversity provided support in the form of an Athena Swan Midcareer Lecturer Research Capacity Building Grant awarded to AG, which was used for publication costs. We thank the anonymous referee for their insightful feedback which enhanced the quality of our manuscript.
\end{acknowledgments}

%
\vspace{5mm}
Simulation data is made available through zenodo \citep{zenodo_data}.
\facility{TBL(NARVAL). Data are made available through the PolarBase archive (\citealt{Petit2014}; http://polarbase.irap.omp.eu/). \\
ICHEC Kay HPC. }


\software{\textsc{Swmf} \citep{Gombosi2018};
            \textsc{Matplotlib} \citep{Hunter2007};
            \textsc{Seaborn} \citep{Waskom2021}; 
            \textsc{NumPy} \citep{Harris2020};
            \textsc{Pandas} \citep{Pandas2020}}



\appendix
\section{Velocity Distributions}
\label{app:v_dists}
\begin{figure}
    \centering
    \includegraphics[trim=0 10 100 0, clip, width=\textwidth]{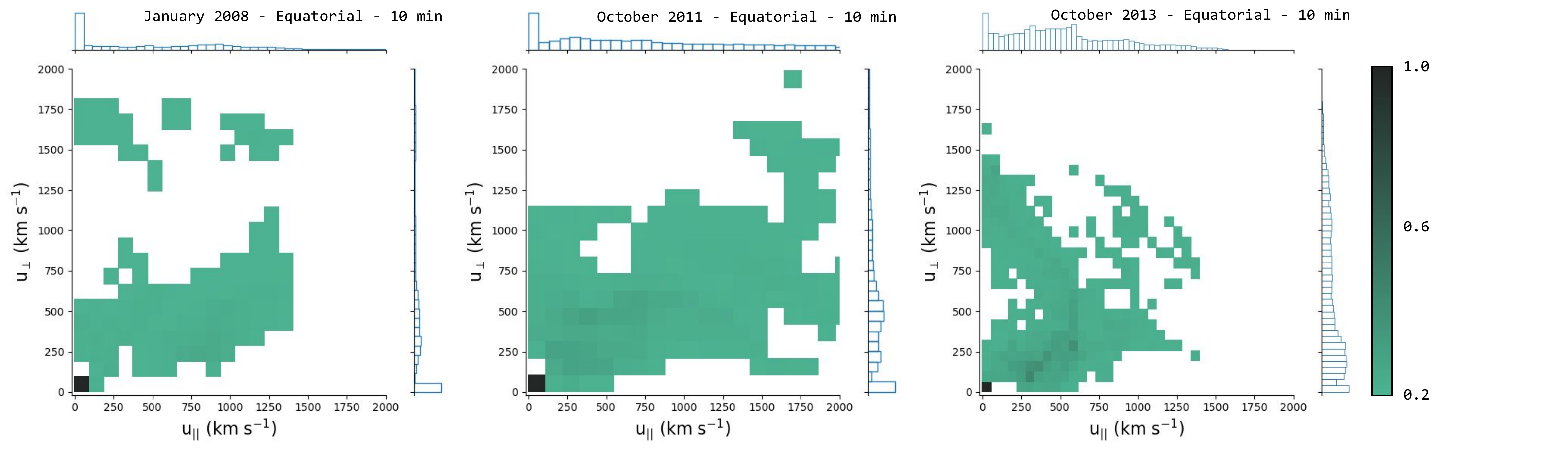}
    \caption{Normalised distributions of parallel and perpendicular velocity magnitudes for each equatorial CME event at 10 minutes post eruption. Peaks near (u$_{||}$,u$_{\perp}$) = (0,0) indicate material that is trapped in coronal loops near the star having not escaped.}
    \label{fig:v_dists}
\end{figure}

We show the magnitude of the orthogonal components of the velocity to the magnetic field within the CME shocks in our simulations in Figure \ref{fig:v_dists}. In particular we show the distribution of each equatorial CME at the 10-minute timestep in our simulations. These components can be derived from our \textsc{awsom} simulations by carrying out a vector projection of the velocity $\mathbf{v}$ onto the magnetic vector $\mathbf{B}$ where
\begin{equation}
    \mathbf{v_{||}} = \frac{\mathbf{B} ( \mathbf{v} \cdot \mathbf{B})}{\left|\left| \mathbf{B} \right|\right| ^{2}}
\end{equation}
\begin{equation}
    \mathbf{v_{\perp}} = \mathbf{v} - \mathbf{v_{||}}
\end{equation}

\begin{figure}
    \centering
    \includegraphics[width=0.85\columnwidth]{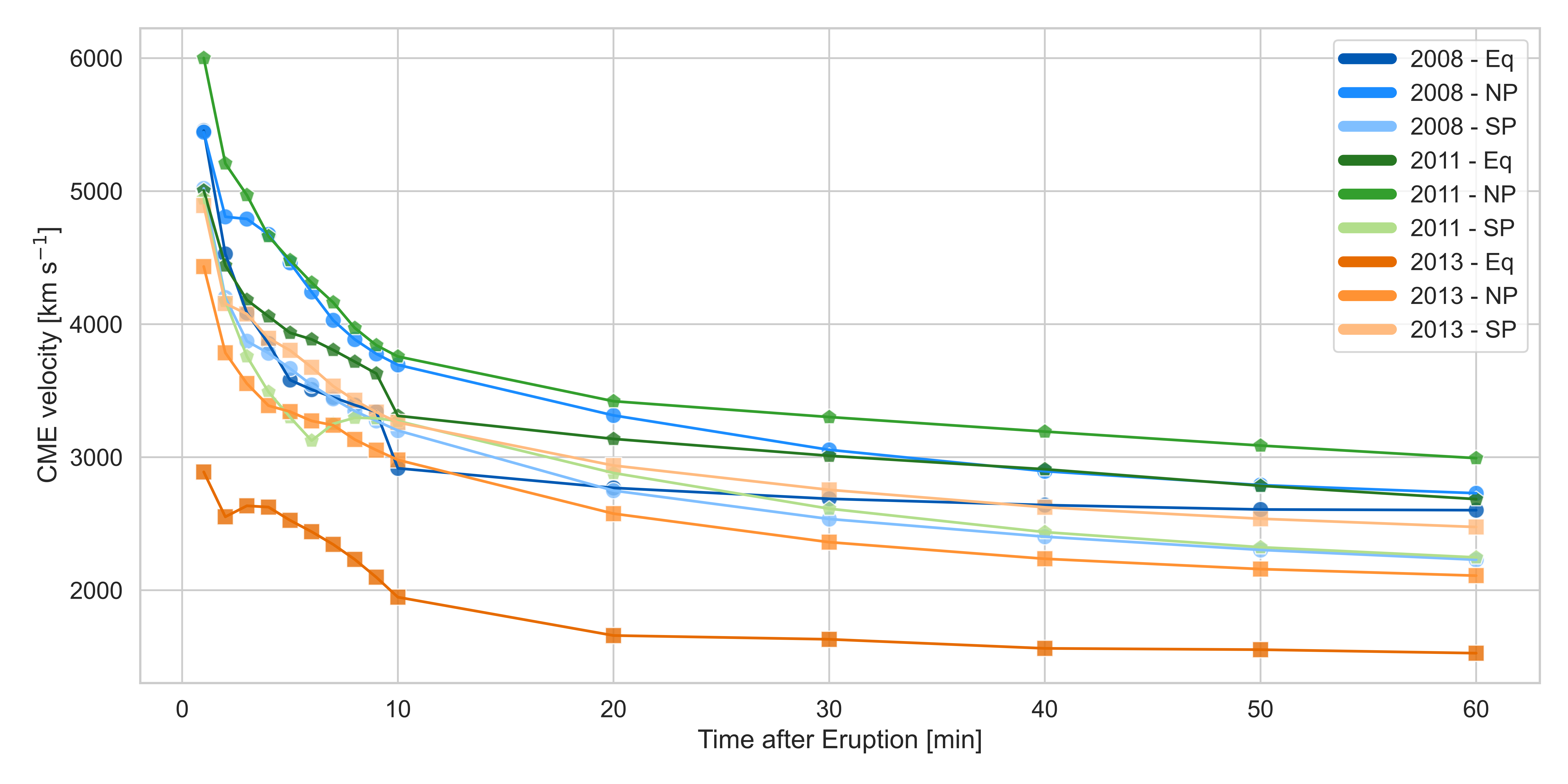}
    \caption{This figure shows the maximum radial velocity of the CME shock. Each colour and marker represents a ZDI epoch containing three CME simulations each. Blue circles (2008), green pentagons (2011), and orange squares (2013).}
    \label{fig:cme_vel}
\end{figure}

We can see that the 2011 epoch has the longest tailed distribution in both $\mathbf{v_{\perp}}$ and $\mathbf{v_{||}}$. These electron distributions could be used as background distributions for electron beams and radiative emission simulations in future work. Figure \ref{fig:cme_vel} shows the maximum velocity of the CME shock over the simulation time for each case. We see that the October 2013 equatorial case has a significantly lower velocity throughout.

\section{Stellar wind simulations}
\begin{figure}[h]
    \centering
    \includegraphics[trim=0 0 0 0, clip, width=0.98\textwidth]{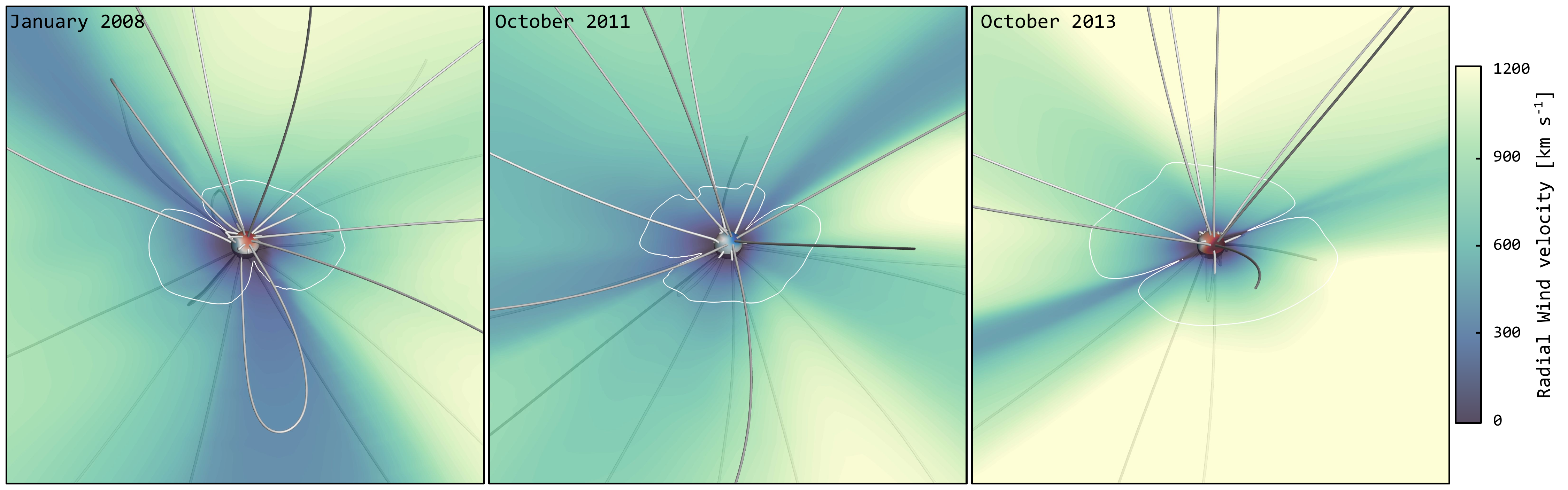}
    \caption{Steady state wind solutions for the three ZDI epochs used in this work. Magnetic streamlines are shown in grey. The colourbar depicts the wind velocity through the equatorial plane. The white line shows where the Alfv\'{e}n surface intersects with the equatorial plane.}
    \label{fig:ss_wind}
\end{figure}

Section \ref{sec:ss-wind} describes the \textsc{AWSOM} wind models that we use as a starting point for our CME simulations. The solutions to these stellar winds are shown here, pre-eruptive event in Figure \ref{fig:ss_wind}. Global wind variables for these solutions are plotted in Figure \ref{fig:mdotjdot}.

\begin{figure}
    \centering
    \includegraphics[width=0.6\columnwidth]{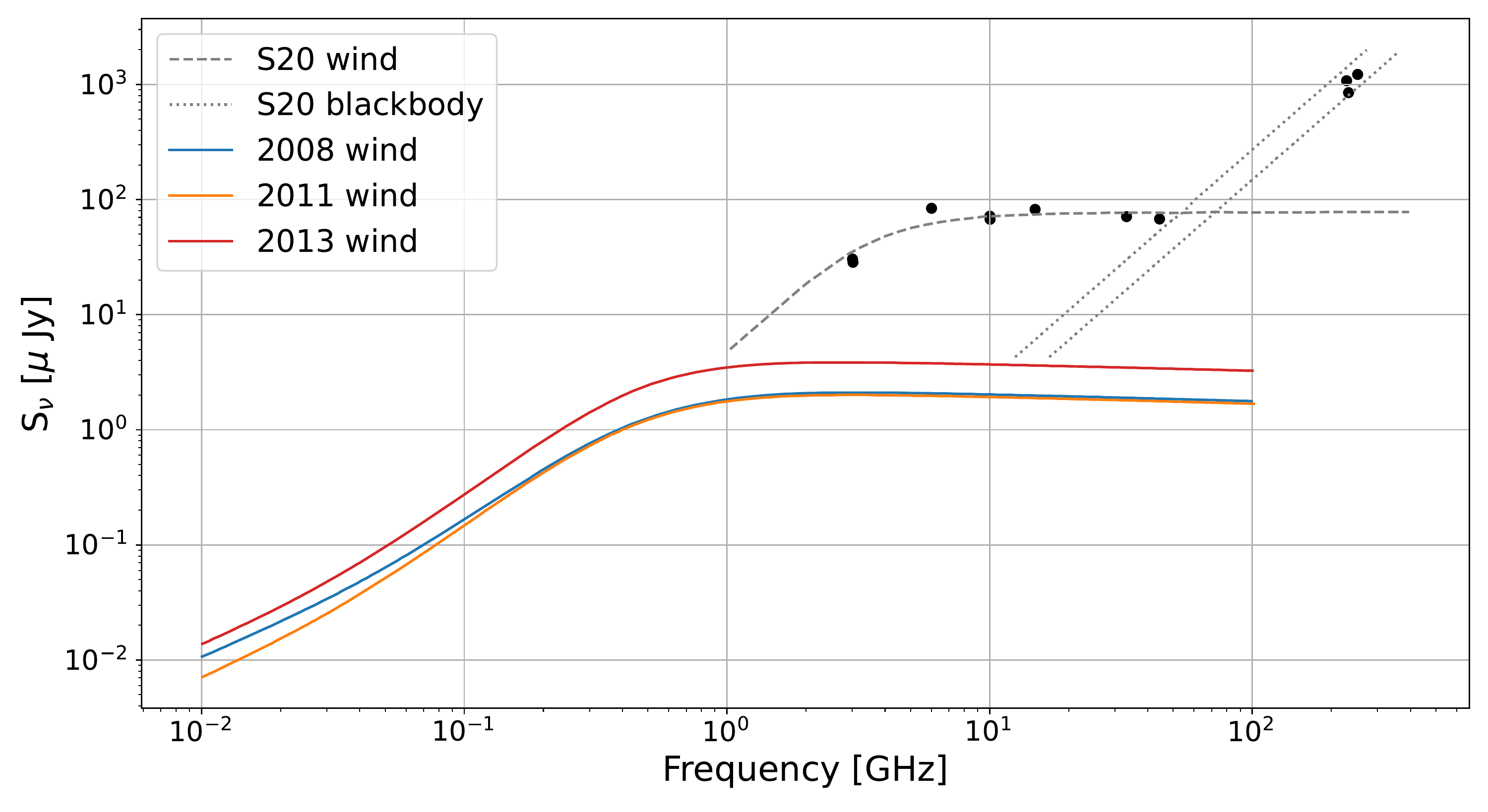}
    \caption{A plot showing the flux density (S$_{\nu}$) against frequency. We compare the thermal radio spectra of our simulated steady state winds to radio observations collated in \citet{Suresh2020}. Black dots show various radio observations with the VLA, ATCA, and ALMA. Grey lines show fits to this data from \citet{Suresh2020}. Simulated wind spectra are shown in blue (2008), orange (2011), and red (2013).}
    \label{fig:wind_spec}
\end{figure}

We compare the thermal radio spectrum calculated from our 3D MHD wind simulations to the radio observations from \citet{MacGregor2015,Booth2017,Lestrade2015,Bastian2018,Rodriguez2019,Suresh2020} in Figure \ref{fig:wind_spec}. We find a much lower flux density for the wind spectra for each epoch simulated here, compared to the 6-20 GHz range of observations (referenced previously and collated in \citealt{Suresh2020}). This supports conclusions drawn by \citet{Bastian2018}, that these radio detections were indeed due to high density active regions. Alternatively these could be chromospheric detections, which is also much denser than the stellar wind, giving a higher flux density. Our simulations are in agreement with these data points in that our wind is optically thin in the same regime. This difference in estimated spectra is not unexpected given the large difference in mass-loss rate from the different wind estimates (i.e. \citealt{Wood2018} vs. \citealt{Suresh2020}).

\section{Shock Definition} \label{app:shocks}
\begin{figure}
    \centering
    \includegraphics[width=0.6\columnwidth]{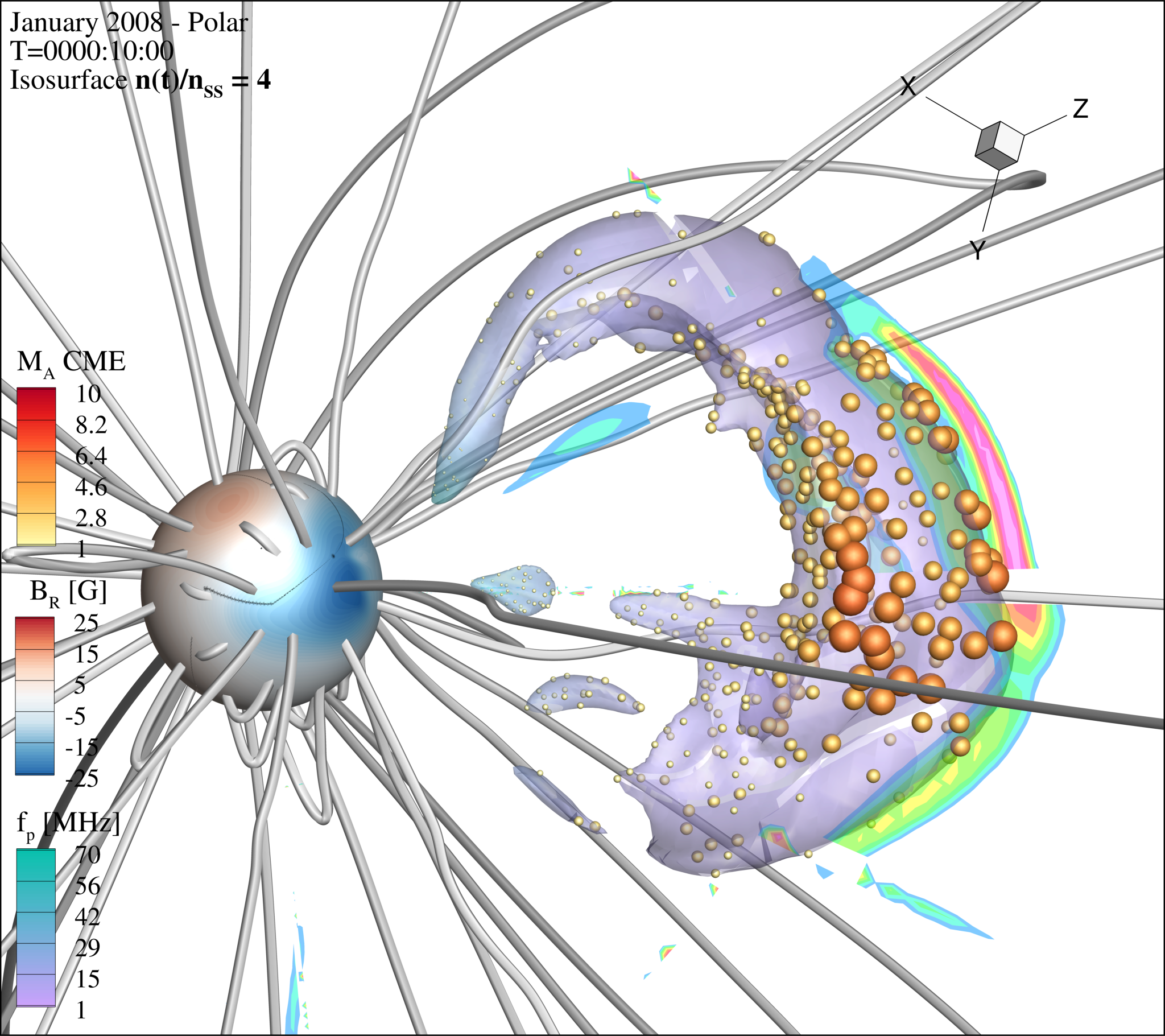}
    \caption{Similar to Figure \ref{fig:j08_eq_cme}, with the addition of the location of largest entropy gradient (from blue to pink) shown as a 2D slice around the previously defined CME isosurface. The displacement between the furthest CME isosurface edge and maximum entropy gradient is 0.2 R$_{\star}$.}
    \label{fig:entropy_shock}
\end{figure}

In this work, we use the edge of the CME to define the shock location and the plasma frequency of the emission. In reality, the shock would exist preceding the CME itself. \citet{Schmidt2013} used the entropy gradient in their simulation to define the shock location and normal. Using this method in our simulation results in the shock front shown in Figure \ref{fig:entropy_shock}. However, we avoided using this method as it resulted in a large number of artefacts incorrectly attributed to the shock, skewing our estimates of the plasma frequency for the shock. We can see that the defined shock contours (blue-pink) using the entropy gradient method surround the delineated CME isosurface. This definition extends $0.2 R_{\star}$ further radially than the CME isosurface, resulting is some lower densities at this point. As a result, our calculated plasma frequencies are upper limits at each timestep in our simulations, and the true value is likely between $\times1-2$ lower in frequency. More robust shock delineation and radio emission simulations are necessary to more accurately describe the type II bursts. 

\bibliography{references}
\bibliographystyle{aasjournal}



\end{document}